\documentclass[12pt]{article}
\usepackage{amsmath}
\usepackage{graphics}
\usepackage{latexsym}

\usepackage [ansinew]{inputenc}
\usepackage {hyperref}
\begin{document}
\vskip 2cm
\title{ Novel Explicit Multi Spin String Solitons in $AdS_5$}

\author{\\
A.L. Larsen${}^{*}$ and
A. Khan${}^{|}$}
\maketitle
\noindent
{\em Physics Department, University of Southern Denmark,
Campusvej 55, 5230 Odense M,
Denmark}

\vskip 8cm
\noindent
$^{*}$Electronic address: all@fysik.sdu.dk\\
$^{|}$Electronic address: patricio@fysik.sdu.dk

\newpage
\begin{abstract}
\baselineskip=1.5em
\hspace*{-6mm} We find new explicit solutions describing closed strings
spinning with equal angular momentum
in two independent
planes in $AdS_5$. These are $2N$-folded strings in the
radial direction
and also winding $M$ times around an angular direction. Thus in spacetime they consist of $2N$
segments. Solutions
fulfilling the closed string
periodicity conditions exist provided $N/M>2$, i.e. the  strings
must be folded at least six times in the radial coordinate.
The strings are spinning, or actually orbiting, 
similarly to solutions found previously
in black hole spacetimes, but unlike the one-spin solutions in $AdS$ which
spin around
their center. For long strings we recover the logaritmic scaling relation
between energy and spin  known from
the one-spin case, but different from other known two-spin cases.
\end{abstract}

\newpage
\newpage

\section{Introduction and Results}
\setcounter{equation}{0}
In connection with the conjectured
duality \cite{maldacena,gubser,witten} between super string theory on
$AdS_5\times S^5$ and
${\cal N}=4$ SU(N)
super Yang-Mills theory in Minkowski space, there has been a lot of
interest in string solitons in $AdS_5$, $S^5$,
$AdS_5\times S^5$ and other related backgrounds. Rigidly rotating strings in $AdS$ \cite{inigo}
(as well as circular pulsating strings \cite{de vega})
were considered long time ago, but
the direct connection  with gauge theory was only discovered quite recently
\cite{klebanov}. Namely, it was
noticed  that the
subleading term in the scaling relation between energy $E$ and spin $S$, 
for the rigidly rotating strings, is
logarithmic
\begin{equation}
 E-S\sim \ln(S)
\label{eq1.1}
\end{equation}
similarly to results found for certain Yang-Mills operators back in the
early days of QCD
\cite{gross,georgi,floratos, korchemsky, dolan}. The paper \cite{klebanov}
initiated a whole industri of
finding new string solitons, and in some cases also the corresponding
Yang-Mills operators. See for instance
\cite{tseytlin,tseytlin2,russo,armoni,
mandal,minahan,
barbon,axenides,alishabiba,buchel,rashkov,ryang,bozhilov,bozhilov2,patricio}
for various string configurations.

More recently it was discovered that multi spin solutions could also be
constructed easily \cite{tseytlin3}. Among others, a relatively
simple solution was found  in $AdS_5$ describing a string which is located
at a point in the radial direction, winding around an
angular direction and spinning with equal angular momentum in two
independent planes.  In the long string limit it was shown that
\begin{equation}
 E-2S\sim S^{1/3}
\label{eq1.2}
 \end{equation}
differing substantially from (\ref{eq1.1}). This is not at all a problem though, since
the subleading terms are well-known to depend
on the particular string configuration.  Multi spin solitons have been
further investigated in a number of papers including
\cite{patricio,engquist,beisert,tseytlin4,tseytlin5,tseytlin6,tseytlin7,tseytlin8,stefan,kim} (for a
review see \cite{tseytlin9}).   In particular, in \cite{tseytlin6} it was
shown that the solutions
to the string equations of motion can be classified  and related to
solutions of the integrable Neumann system. However,
multi spin solutions were mostly considered in $S^5$
in the abovementioned papers.  Even though $AdS_5$ and $S^5 $
geometrically only differ by a couple of signs, the scaling
relations between energy and spin for string solitons  are completely
different. So, much more work needs to be done for multi spin
solutions in $AdS_5$.

While many explicit multi spin solutions in $S^5$ have been constructed
\cite{tseytlin3,tseytlin4,tseytlin5,tseytlin6,tseytlin7,tseytlin8,tseytlin9}, 
it seems that there are essentially
still only two explicit
multi spin solutions known in $AdS_5$, namely the one-parameter family
\cite{tseytlin3}  leading to (\ref{eq1.2}) and
its generalisation to arbitrary unequal angular momenta \cite{tseytlin8,tseytlin9},
which gives a somewhat similar
(although non-regular) scaling relation.

In the present paper we
construct another one-parameter family of multi spin solutions in $AdS_5$.
Our solutions describe  $2N$-folded 
strings in the radial direction
and also winding $M$ times around an angular direction. They are spinning
with equal angular momentum in
two independent planes. For some reason, for which we have no physical
explanation,
the closed string
periodicity conditions can only be fulfilled provided $N/M>2$, i.e., the 
strings must be folded at least six times in the radial coordinate. But it must be stressed that the 
strings are not folded in spacetime; they consist of $2N$ segments.

Moreover, our strings are quite different from the usual straight strings
spinning in  $AdS$ \cite{inigo}, since
they are not spinning around their center. Instead, the whole string is
spinning, or actually orbiting,  around origo similarly to solutions previously found
in black hole spacetimes \cite{inigo,armoni,kar,hendy}. There is of course no central 
attractive body in this case; instead, the orbits are stabilised by the string winding around
an angular direction.  

For long strings we find that
\begin{equation}
 E-2S\sim \ln(S)
\label{eq1.3}
\end{equation}
similarly to (\ref{eq1.1}) except for the obvious factor 2, but completely
different from (\ref{eq1.2}).
 Thus we recover the remarkable logaritmic
scaling relation from the one-spin case \cite{klebanov}.  It must be stressed that our
solutions are
"orthogonal" to the solutions of \cite{tseytlin3} in the sense that there
is only one common solution
 in the two families,
and this particular solution is not a long string.

The paper is organised as follows.  In Section 2, we present our ansatz and solve the resulting
equations of motion explicitly. We also obtain the periodicity conditions in compact form. In Section 3, we solve 
the periodicity conditions for short and long strings, and analyse the corresponding string configurations. We also
comment on the relation to previously known results. In Section 4, we compute the energy $E$ and the two spins
$S_1=S_2\equiv S$, and derive the logaritmic scaling  relation (\ref{eq1.3}) in the limit of long strings.
Finally in Section 5,
we present our conclusions as well as some comments about stability.  
\section{ Two-Spin Solutions  in $AdS_ 5$  }
\setcounter{equation}{0}
We take the $AdS_5$  line-element  in the form
    \begin{eqnarray}
 ds^2&=&-(1+H^2r^2)dt^2+\frac{dr^2}{1+H^2r^2}+r^2(d\beta^2+\sin^2\beta
d\phi^2+\cos^2\beta d\tilde{\phi}^2 )
    \end{eqnarray}
where  $H^{-1}$ is the scale of $AdS_5$. The 't
Hooft coupling in this notation is
$\lambda =(H^2\alpha')^{-2}$.

The string which is extended in the $r$ and $\beta$ directions and spinning
with equal spin in the
$\phi$ and $\tilde{\phi}$ directions, is obtained by the
 ansatz
\begin{equation}
t=c_0\tau,\ \ \ r=r(\sigma), \ \ \ \beta=\beta(\sigma ),\ \ \
\phi=\omega\tau, \ \ \     \tilde{\phi}=\omega\tau
\end{equation}
where $c_0$ and $\omega $ are arbitrary constants.
 Then the $r$ and
$\beta $ equations become
\begin{equation}
r''-\frac{{r'}^2H^2r}{1+H^2r^2}-{\beta'}^2(1+H^2r^2)r-H^2c_0^2(1+H^2r^2)r+\omega
^2(1+H^2r^2)r=0
\end{equation}
\begin{equation}
\beta''+\frac{2r'\beta'}{r}=0
\end{equation}
while the non-trivial conformal gauge constraint  is
\begin{equation}
\frac{{r'}^2}{1+H^2r^2}+r^2{\beta'}^2-c_0^2(1+H^2r^2)+r^2\omega^2=0
\label{2.5}
\end{equation}
    The system is reduced to first order form
\begin{equation}
\beta'=\frac{k}{r^2}
\label{eq2.6}
    \end{equation}
   \begin{equation}
\frac{{r'}^ 2}{1+H^2r^2}+\frac{k^2}{r^2}+r^2\omega ^2-c_0^2(1+H^2r^2) =0
\label{eq2.7}
    \end{equation}
where $k$ is an integration constant. These equations are identical to
those considered in \cite{tseytlin3} (in the parametrisation $Hr=\sinh\rho$),
but only
constant $r$ solutions were found there. Similar equations and their
generalisations were also considered
in
 \cite{tseytlin8,tseytlin9}, but again only constant $r$ solutions were
explicitly constructed.
 We shall now show that solutions with $r=r(\sigma )$ exist as well,
provided that the
string is folded at least six times in the radial direction.

First concentrate on the $r$ equation (\ref{eq2.7}). In the non-negative
dimensionless coordinate $y\equiv H^2r^2$, it can be
written
     \begin{equation}
{{y'}^ 2}=4(H^2c_0^2-\omega^2)y^3+4(2H^2c_0^2-\omega ^2)y^
2+4(H^2c_0^2-H^4k^2)y-4H^4{k^2}
    \end{equation}
    If $H^2c_0^2-\omega ^2=0$, the solution is
    \begin{equation}
y=\frac{-(c_0^2-H^2k^2)}{2c_0^2}\pm
\frac{(c_0^2+H^2k^2)}{2c_0^2}\cosh(2Hc_0\sigma)
    \end{equation}
    but it must be discarded since it is not periodic.
    If $H^2c_0^2-\omega ^2\neq 0$,
     we notice that the right hand side can be factorized
       \begin{eqnarray}
{y'}
^2&=&   4\left(H^2c_0^2-\omega^2\right)\left(y+1\right)\nonumber\\
&&\left(y-\left(\frac{H^2(-c_0^2+\sqrt{c_0
^4+4k^2(H^2c_0^2-\omega^2)}}{2(H^2c_0^
2-\omega^2)}\right)\right)\nonumber\\
&&\left(y-\left(\frac{H^2(-c_0^2-\sqrt{c_0^4+4k^2(H^2c_0^2-\omega^2)}}{2(H^2c_0^
2-\omega^2)}\right)\right)
 \end{eqnarray}
 To get periodic solutions, we are interested in the case with two
non-negative roots, so we get the
conditions
        \begin{equation}
H^2c_0^2-\omega ^2<0
 \label{eq2.11}   \end{equation}
      \begin{equation}
c_0^4+4k^2(H^2c_0^2-\omega ^2)\geq 0
\label{eq2.12}    \end{equation}
   Now define the function $P$
      \begin{equation}
    y=\frac{P}{H^2c_0^2-\omega ^2 }-\frac{2H^2c_0^2-\omega
^2}{3(H^2c_0^2-\omega^2)}
        \end{equation}
 such that
      \begin{equation}
      {P'}^2=4P^3-g_2P-g_3=4(P-e_1)(P-e_2)(P-e_3)
  \label{eq2.14}   \end{equation}
   where the roots $(e_1>e_2\geq e_3)$ are
    \begin{equation}
e_1=\frac{1}{3}(2\omega^2-H^2c_0^2)
    \end{equation} \begin{equation}
e_2=\frac{1}{6}H^2c_0^2-\frac{1}{3}\omega^2+\frac{1}{2}H^2
\sqrt{c_0^4+4k^2(H^2c_0^2-\omega^2)}
    \end{equation} \begin{equation}
e_3=\frac{1}{6}H^2c_0^2-\frac{1}{3}\omega^2-\frac{1}{2}H^2
\sqrt{c_0^4+4k^2(H^2c_0^2-\omega^2)}
    \end{equation}
    and the invariants are $g_2=2(e_1^2+e_2^2+e_3^2)$ and
$g_3=4e_1e_2e_3$. Notice also that   $\Delta \equiv g_2^3-27g_3^2\geq 0$
(using (\ref{eq2.11}), (\ref{eq2.12})).

 Equation (\ref{eq2.14}) is the Weierstrass equation with
solution
      \begin{equation}
P(\sigma )={\wp}(\sigma +a)
    \end{equation}
    where $\wp$ is the doubly-periodic Weierstrass function and  $a$ is a
complex  integration constant
\cite{abramowitz}.
 To get a real non-singular solution, we take  $a$ to be half the imaginary
period. Then the Weierstrass function reduces
 to a Jacobi elliptic function \cite{abramowitz}.        Before writing
down the explicit solution,
 it is convenient to trade the parameters $(c_0,\omega, k)$ for
 \begin{equation}
b=e_2-e_3=    H^2\sqrt{c_0^4+4k^2(H^2c_0^2-\omega^2)}
\end{equation}
\begin{equation}
m=\frac{e_2-e_3}{e_1-e_3}=\frac{2H^2\sqrt{c_0^4+4k^2(H^2c_0^2-\omega^2)}}{2\omega^2-H^2c_0^2
+H^2\sqrt{c_0^4+4k^2(H^2c_0^2-\omega^2)}}
\end{equation}
\begin{equation}
n=\frac{2\sqrt{c_0^4+4k^2(H^2c_0^2-\omega^2)}}{c_0^2+\sqrt{c_0^4+4k^2(H^2c_0^2-\omega^2)}}
\end{equation}
such that by eqs.(\ref{eq2.11})-(\ref{eq2.12})
\begin{equation}
 b\geq 0,\ \ \ \ 0\leq m\leq n\leq 1
    \end{equation}
  In this parametrisation the solution to (\ref{eq2.7}) is
    \begin{equation}
H^2r^2(\sigma )=          \frac{m}{n-m}\left(1-n
{\mbox{sn}}^2\left(\sqrt{\frac{b}{m}}   \sigma|m\right)\right)
 \label{eq2.23}     \end{equation}
so that $Hr(\sigma )$ oscillates between the non-negative values $Hr_{max}$
and $Hr_{min}$, where
\begin{equation}
Hr_{max}=\sqrt{\frac{m}{n-m}},\ \ \ \  Hr_{min}=\sqrt{\frac{m(1-n)}{n-m}}
\end{equation}
This means that the whole string is spinning, or actually orbiting,  around origo. It is
quite similar to straight strings spinning around black holes
\cite{inigo,armoni,kar,hendy}, but unlike
one-spin strings in $AdS$ which spin around their center \cite{inigo}.

The solution ({\ref{eq2.23}) must be supplemented with the closed string periodicity
condition $r(\sigma+2\pi )=r(\sigma )$.
For a $2N$-folded closed string ($N$ positive integer), this condition becomes
\begin{equation}
\frac{\sqrt{m}K(m)}{\pi\sqrt{b}}    =\frac{1}{N}
 \label{eq2.25}  \end{equation}
     which trivially gives $b$ in terms of $m$.  This condition of course
drops out for $n=0$ (which implies $b=m=0$), corresponding
     to strings  which are  pointlike in the radial direction
\cite{tseytlin3}. Here we shall only consider strings which are extended in the
radial direction.

     Now return to the $\beta$ equation (\ref{eq2.6}). It is trivially integrated to
\begin{eqnarray}
\beta(\sigma )&=&\sqrt{\frac{b(n-m)(1-n)}{nm}}\int_0^\sigma\frac{dx}{1-n
{\mbox{sn}}^2\left(\sqrt{\frac{b}{m}}   x|m\right)}
\nonumber \\
&=& \sqrt{\frac{(n-m)(1-n)}{n}}\Pi\left(n;\sqrt{\frac{b}{m}}\sigma|m\right)
\end{eqnarray}
  Since $\beta $ is an
angular coordinate, we impose the quasi-periodicity  condition
    $\beta (\sigma +2\pi)=2M\pi+\beta (\sigma )$, where $M$ is another
positive  integer which plays the role of a winding number.
    This condition translates into
                 \begin{equation}
\frac{\sqrt{(1-n)(n-m)}\Pi(n;m)}{\pi\sqrt{n}} =\frac{M}{N}
 \label{eq2.27}     \end{equation}
         where (\ref{eq2.25}) was also used. This equation determines $n$ in terms
of $m$, thus the only free parameter (for fixed $N$, $M$)
      now is $m$.
      \section{Analysis of the Solutions}
  \setcounter{equation}{0}
The main problem  when analysing the solutions of the previous section, is
that (\ref{eq2.27}) is trancendental.
Numerical analysis shows that the left hand side is less than 1/2 (it
equals 1/2 only for $m=0$, corresponding
to strings which are pointlike at $r=0$). Thus, to get non-trivial
solutions, we need $N>2M$ which means that
the string must be folded at least six times. In that case we get a
one-parameter family of solutions with
arbitrary extension in the radial direction.

 For short and long strings, we can solve (\ref{eq2.27}) analytically and
confirm the numerical results. First
consider short strings.
 From  (\ref{eq2.23}) follows that short strings correspond to
 $m\approx 0$, $n\approx 0$ (with $m\leq n$).  Using that $\Pi(n;m)\approx
\pi/2$ in this case, we get from
(\ref{eq2.27})
 \begin{equation}
\frac{\sqrt{n-m}}{2\sqrt{n}}\approx \frac{M}{N}
\end{equation}
such that
 \begin{equation}
 \frac{n}{m}\approx \frac{N^2 }{N^2-4M^2},\ \ \ \ {\mbox{short strings}}
\label{eq3.2}    \end{equation}
which implies $N>2M$.
 For such short strings, in the limit where $r$ is almost a constant, we get
\begin{equation}
Hr\approx\sqrt{\frac{N^2-4M^2}{4 M^2}}
\end{equation}
\begin{equation}
\beta\approx M\sigma
\label{eq3.4}    \end{equation}
thus the string is located at a fixed finite  value of $r$ and simply
winding around  in the $\beta $ direction. This is a special
case of the solutions found in  \cite{tseytlin3}.  To get all the solutions
of \cite{tseytlin3}, one should discard
the periodicity condition, as discussed after (\ref{eq2.25}). But notice that
our notion of short means short in the radial extension,
which is different from the notion of short used elsewhere.

Now consider long strings, which correspond to $n\approx 1, \ m\approx 1$
(with $m\leq n$). Here we use the
approximation
 \begin{equation}
 \Pi(n;m) \approx\frac{\pi }{2}  \sqrt{\frac{1}{(1-n)(n-m)}}
 \left(1-\frac{2}{\pi}\sin^{-1 }\sqrt{\frac{1-n}{1-m}}\right)
\end{equation}
Then (\ref{eq2.27}) gives
\begin{equation}
\frac{M}{N}     \approx \frac{1}{2}-\frac{1}{\pi}\sin^{-1
}\sqrt{\frac{1-n}{1-m }}
\end{equation}
such that
    \begin{equation}
\sqrt{\frac{1-n}{1- m}}\approx\cos\left(\frac{M}{N} \pi\right),\ \ \ \
{\mbox{long strings}}
\label{eq3.7}    \end{equation}
which again implies $N>2M$.
This leads to the following expression for $Hr_{min}$
\begin{eqnarray}
Hr_{min}\approx{\mbox{cotan}}\left(\frac{M}{N}\pi\right)
\end{eqnarray}
Thus the long strings extend outwards  from this fixed value.
We observe that the long strings come closer to origo than the short strings.
\section{Energy and Spin}
   \setcounter{equation}{0}
The conserved energy is given by
 \begin{eqnarray}
 E=\frac{c_0}{2\pi\alpha'}\int_0^{2\pi}(1+H^2r^2(\sigma ))d\sigma
 \end{eqnarray}
 In our parametrisation, this can be expressed in terms of a complete
elliptic integral as follows
 \begin{eqnarray}
E
&=&\frac{N\sqrt{(2-n)nm}}{\pi H\alpha'(n-m)}E(m)
\end{eqnarray}
Similarly, the conserved spins are
\begin{eqnarray}
S_1=\frac{\omega}{2\pi\alpha'}\int_0^{2\pi}r^2(\sigma )\cos^2\beta (\sigma
)d\sigma, \ \ \ \
  S_2=\frac{\omega}{2\pi\alpha'}\int_0^{2\pi}r^2(\sigma )\sin^2\beta
(\sigma )d\sigma
  \end{eqnarray}
  such that  \cite{tseytlin6}
 \begin{eqnarray}
\frac{E}{H^2c_0}-\frac{S_1}{\omega}-\frac{S_2}{\omega}=\frac{1}{H^2\alpha'}
 \end{eqnarray}
 In our case, $S_1=S_2\equiv S=(S_1+S_2)/2$, which   in terms of complete
elliptic integrals gives
\begin{eqnarray}
S
&=&\frac{N\sqrt{n+m-nm}}{2\pi H^2\alpha'(n-m)\sqrt{n}} ((m-n)K(m)+nE(m))
\end{eqnarray}
Further simplifications are obtained for short and long strings, respectively.
For short strings, using (\ref{eq3.2}), we arrive at the following finite
expressions for
    the energy and spin
       \begin{eqnarray}
E&\approx&\frac{N^2\sqrt{N^2-4M^2}}{4\sqrt{2}H\alpha'M^2}\\
S&\approx&\frac{(N^2-4M^2)\sqrt{N^2-2M^2}}{8\sqrt{2}\alpha'H^2M^2} \label{eq4.7}
    \end{eqnarray}
But recall that these are just the minimal values within our
family of strings. As explained after (\ref{eq3.4}),
   these strings are not short in the usual sense.
	In fact, we dont have solutions with $E \approx 0 $  and $S\approx 0$
	within our family of strings. 

       For the long strings we can eliminate $n$ using eq.(3.7) and  get
    \begin{eqnarray}
E&\approx&\frac{N}{\pi H \alpha'\sin^2\left(\frac{M}{N}\pi\right)(1-m)}
\left(1+\frac{1-m}{4}\log\frac{16}{1-m}\right) \\
S&\approx&\frac{N}{2\pi\alpha'H^2\sin^2\left(\frac{M}{N}\pi\right)(1-m)}
\left(1-\left(2\sin^2\left(\frac{M}{N}\pi\right)-1\right)\frac{1-m}{4}\log\frac{16}{1-m}\right)
   \end{eqnarray} 
which leads to
   \begin{equation}
E/H-2S\approx\frac{N}{2\pi H^2 \alpha'}\log (2\pi H^2\alpha' S  )
\end{equation}up to an unimportant additive constant.

Thus, we recover the remarkable and celebrated logaritmic  scaling relation known from the
one-spin case \cite{klebanov}. This should be
contrasted with the two-spin solutions  considered in  \cite{tseytlin3},
where it was found that (in our notation)
\begin{eqnarray}
E/H-2 S\approx \frac{3(H^2\alpha'S)^{1/3}}{2^{4/3}H^2\alpha'}
\end{eqnarray}  
\section{ Concluding Remarks}
\setcounter{equation}{0}
In conclusion, we have constructed a new one-parameter family of multi spin string solitons 
in $AdS_5$.  Contrary to the well-known one-spin solutions in  $AdS$ \cite{inigo}, our 
solutions are orbiting around origo. More importantly, contrary to other explicitly known
multi spin solutions \cite{tseytlin3,tseytlin8}, our solutions give the logaritmic scaling 
relation between energy and total spin $E-2S\sim \ln(S)$. This suggests that the corresponding 
Yang-Mills operator can be constructed along the lines of \cite{klebanov}.

Our solutions can be generalised straightforwardly  in various ways to solutions in $AdS_5\times S^5$.
Either by adding one or more R-charges \cite{tseytlin3} or by adding pulsation \cite{patricio}.

In order to test the stability of our solutions, it would be interesting to  consider 
lineralised perturbations around them.
 Unfortunately,  it turns out to be extremely
complicated   because of the highly  
non-trivial sigma-dependent $r(\sigma)$ and $\beta (\sigma )$.
As a  result, we will get sigma-dependent coefficients in the 3 coupled  
equations for the physical perturbations, which
should be compared with the case considered in \cite{tseytlin3} where the coefficients were constant. 
At the moment, we have no solution to this problem. 	 However, we can test our solutions in
the limit where they reduce to a solution of \cite{tseytlin3}. Namely our short strings (which are not short 
in the sense of \cite{tseytlin3}), where the spin 
is given by (\ref{eq4.7}). If we take the simplest allowed case   ($N=3,\ M=1$),
we get $S=5\sqrt{7\lambda /2}/8 \approx 1.17\sqrt{\lambda}$,  which is far below the set-in of
instability found in \cite{tseytlin3}, thus ensuring us of a stable solution.
		 
\vskip 24pt
\hspace*{-6mm}{\bf Acknowledgements}:\\
We would like to thank  N. Sibani for his help in the preparation of this
paper.

\end{document}